\documentclass[pre,showpacs,aps,twocolumn]{revtex4}
\usepackage{epsfig}
\usepackage{bm}
\usepackage{amssymb} 

\renewcommand{\it}[1]{\textit{#1}}
\newcommand{\Onecol} {\begin{widetext} \onecolumngrid} %% 2 -> 1
\newcommand{\Twocol} {\end{widetext} \twocolumngrid} %% 1 -> 2
\newcommand{\be}{\begin{equation}}
\newcommand{\ba}{\begin{array}}
\newcommand{\bea}{\begin{eqnarray}}
\newcommand{\bfi}{\begin{figure}}
\newcommand{\ee}{\end{equation}}
\newcommand{\ea}{\end{array}}
\newcommand{\eea}{\end{eqnarray}}
\newcommand{\efi}{\end{figure}}

\newcommand{\ra}{\right\rangle}
\newcommand{\la}{\left\langle}
\newcommand{\G}{{\cal G}}
\newcommand{\x}{{\bm x}}
\newcommand{\dd}{{\mbox{d}}}

\begin{document} 

\bibliographystyle{prsty}

\title{Mesoscopic two-phase model for describing apparent slip in  micro-channel flows}
\author{R. Benzi$^1$, L. Biferale$^1$, M. Sbragaglia$^1$, S. Succi$^2$ and F.
Toschi$^{2,3}$}
\affiliation{$^1$ Dipartimento di Fisica and INFN, Universit\`a di Tor
 Vergata, via della Ricerca Scientifica 1, 00133, Roma, Italy. \\
$^2$ Istituto Applicazioni Calcolo, CNR, Viale del Policlinico 137, 00161 Roma, Italy.\\
$^3$ INFN, Via del Paradiso 12, 44100 Ferrara, Italy.
} 


\begin{abstract} 
The phenomenon of apparent slip in micro-channel flows is analyzed by
means of a two-phase mesoscopic lattice Boltzmann model including
non-ideal fluid-fluid and fluid-wall interactins.  The
weakly-inhomogeneous limit of this model is solved analytically.
 The present mesoscopic approach permits to access much
larger scales than molecular dynamics, and comparable with those
attained by continuum methods. However, at variance with the continuum
approach, the existence of a gas layer near the wall does not need to
be postulated a priori, but emerges naturally from the underlying
non-ideal mesoscopic dynamics.  It is therefore argued that a
mesoscopic Lattice Boltzmann approach with non-ideal fluid-fluid and
fluid-wall interactions might achieve an optimal compromise between
physical realism and computational efficiency for the study of channel
micro-flows.
\vskip 0.2cm 
\end{abstract} 

\maketitle 

%%%%%%%%%%%%%%%%%%%%%%%%%%%%%%%%%%%%%%%%%%%%%%%%%%%%%%%%%%%%%%%%%%%%%%%%%%%%
The microscopic physics underlying fluid/solid interactions is fairly
rich and complex, for it depends on specific details of molecular
interactions as well as on the micro-geometrical details of the
boundary.  However, on a macroscopic scale, these details can often be
safely ignored by assuming that the net effect of surface interactions
is simply to prevent any relative motion between the solid walls
and the fluid elements next to them.  This is the so-called ``no-slip''
boundary condition, which forms the basis of mathematical treatments
of bounded flows as continuum media \cite{Gold}.  No-slip
boundary conditions are extremely successful in describing a huge
class of viscous flows. Yet, the evidence is that certain classes of
viscous flows {\em do} slip on the wall. Recent advances in
microfluidics experiments \cite{MEMS,tabebook}, as well as numerical
investigations \cite{MD2,MD6,MD3,troian,barrat}, have identified the conditions
which seem to underlie the validity of the no-slip assumption.  Namely:
(i) single-phase flow; (ii) wetted surfaces and (iii) low levels of
shear rates.  Under such conditions, careful experiments have shown
that fluid comes to rest within a few molecular diameters from the
surface \cite{Chan,Israelachvili,Klein,Raviv}.  Conditions (i-iii) are
not exhaustive, though.  For instance, partial slips of simple
(Newtonian) flows, such as alkanes and water, is predicted by an
increasing number of experiments \cite{choi,cheng,pit,tab} and simulations
\cite{MD2,MD3,MD6,troian,barrat} (see  \cite{brenner} for a review on
experiments and numerics).
Under this state of affairs, there appears to be a great need to
provide a convincing, and possibly general, theoretical picture for
the onset of slip motion.  Among others, an increasingly popular
explanation is that the flowing fluid would develop a lighter (less
dense) phase and dynamically segregate it in the form of a thin film
sticking to the wall \cite{tretheway,degennes}.  This thin film would
then provide a ``gliding'' surface for the bulk fluid which would slip
on it without ever coming in contact with the solid wall.  This gives
rise to the so-called {\em apparent slip} phenomenon, that is, the
extrapolated bulk flow speed would vanish far-out away from the wall,
even though the actual flow speed in the film does vanish exactly at
the wall location.  This film-picture is very appealing, but still in
great need of theoretical clarification. In particular, the underlying
mechanisms of film formation are still under question: are they
generic or detail-driven?\\ 
In this paper we shall propose that film
formation is a {\em generic} phenomenon, which can be captured by a
one-parameter mesoscopic approach, lying in-between the microscopic
(atomistic) and macroscopic (continuum) levels.  The mesoscopic
approach is based on a minimal (lattice) Boltzmann equation, (LBE)
\cite{Saurobook,gladrow,BSV}, including non-ideal interactions
\cite{Shanchen1,BE2phase3,BE2phase4,BE2phase5,KWOK,KWOK2}, which
can drive dynamic phase transitions. The only free parameter in the
LBE is the strength of these non-ideal (potential energy)
interactions.
Hopefully, the present mesoscopic approach provides an optimal
compromise between the need of including complex physics
(phase-transition) not easily captured by a continuum approach, and
the need of accessing experimentally relevant space-time scales which
are out of reach to microscopic Molecular Dynamics (MD)
 simulations \cite{MD2,MD3,troian,barrat}.
In particular, at variance with the macroscopic approach, the gas film
does not need to be postulated a-priori, but emerges dynamically from
the underlying mesoscopic description, by progressive switching of
potential interactions.
%Conversely, by switching these interactions off the
%present mesoscopic approach goes smoothly into a non-homogeneus fluid
%limit, which can be shown to provide a coherent picture in good
%agreement with experimental and numerical observations, in particular
%a correct prediction of the slip length as a function of the film
%thickness.
One major advantage of this formulation is that it allows to develop a
simple and straightforward analytical interpretation of the results as
well as of the effective slip length arising in the flow.  This
interpretation is based on the macroscopic limit of the model which
can be achieved by a standard Chapman-Enskog expansion.\\ The lattice
Boltzmann model used in this paper to describe multiple phases has
been developed in \cite{Shanchen1}.  Since
this model is well documented in the literature, here we shall provide
only the basic facts behind it.  We recall that the model is a
minimal discrete version of the Boltzmann equation, and reads as
follows: \be
f_{l}(\bm{x}+\bm{c}_{l},t+1)-f_l(\bm{x},t)=-\frac{1}{\tau}\left(
f_{l}(\bm{x},t)-f_{l}^{(eq)}(\bm{x},t) \right)
\label{1}
\ee where $f_l(\bm{x},t)$ is the probability density function
associated to a mesoscopic velocity $\bm{c}_{l}$ and where $\tau$ is a
mean collision time and $f^{(eq)}_{l}(\bm{x},t)$ the equilibrium
distribution that corresponds to the Maxwellian distribution in the
fully continuum limit.  The bulk interparticle interaction is
proportional to a free parameter, $\G_b$, entering  the
balance equation for the momentum change: 
\be \frac{\dd (\rho
\bm{u})}{\dd t}={\bm F} \equiv {\cal G}_{b} \sum_l w_{l}
\Psi\left[\rho(\x)\right] \Psi\left[\rho(\x+{\bm c}_l)\right] {\bm
c}_l
\label{2}
\ee being $w_{l}$ the equilibrium weights and $\Psi$ the potential
function which describes the fluid-fluid interaction triggered by
density variation.  
%The Shan-Chen
%model then reads as a normal LBE where in the equilibrium distribution
%the expression for the momentum is replaced by a corrected expression:
%\be \bm{u}' = \bm{u} - \tau \bm{F} = {1\over {\rho({\bm x})}} \left[
%\sum_i \bm{c}_i f_i -\tau \bm{F} \rho({\bm x}) \right].
%\label{3} 
%\ee 
By Taylor expanding eq.(\ref{2}) one recovers, in the hydrodynamical
limit, the equation of motion for a non-ideal fluid with equation of
state $P=c^{2}_{s}(\rho-\frac{1}{2}\G_{b} \Psi^2(\rho))$,
$c_{s}$  being the sound speed velocity. With the choice
$$\Psi(\rho)=1-\exp(-\rho/\rho_{0})$$ with $\rho_{0}=1$ a reference
density, the model supports phase transitions whenever the control
parameter exceeds the critical threshold $\G_{b}>\G^{c}_b$.  In
our case,  $\G^{c}_b=4$ for an averaged density $\la \rho
\ra=\log(2)$.\\ We consider $\G_{b}$ as an external control parameter,
with no need of responding to a self-consistent temperature dynamics.
It has been pointed out \cite{joseph} that the SC model is affected by
spurious currents near the interface due to lack of conservation of
local momentum.  This criticism, however, rests on an ambiguous
interpretation of the fluid velocity in the presence of intermolecular
interactions.  In fact, spurious currents can be shown to disappear
completely once the {\em instantaneous} pre and post-collisional
currents are replaced by a time-average over a collisional time.  This
averaged quantity is readily checked to fulfill both continuity and
momentum conservation equations without leading to any spurious
current \cite{buick}.
%%%%%%%%%%%%%%%%%%%%%%%%%%%%%%%%%%%%%%%%%%%%%%%%%%%%%%%%%%%%%%%%%
%\begin{figure}[h]
%\begin{center}
%\includegraphics[scale=.5]{contact.eps}
%\caption{Quantitative mapping of the parameters $\G_{b}$ and $\G_{w}$
%  onto the more physical concept of ``contact angle''.  At fixed
% $\G_{b}=4.2$ and changing $\G_{w}$ in an appropriate range we
%  can reproduce different contact angles. The results are obtained
%  with  pure bounce-back boundary conditions for the LBE populations. }
%\label{fig1}
%\end{center} 
%\end{figure} 
%%%%%%%%%%%%%%%%%%%%%%%%%%%%%%%%%%%%%%%%%%%%%%%%%%%%%%%%%%%%%%%%%%%%%%%%%%%%%%%%%
Let us now consider the main result of this letter, namely the
critical interplay between the bulk physics and the presence of wall
effects. In fact, 
in order to make contact with experiments and MD simulations, 
it is important to include fluid-wall interactions, and notably a parametric form of
mesoscopic interactions capable of mimicking wettability properties as
described by contact angles between droplets and the solid wall
\cite{degennesrev}. This effect is
 achieved by assuming that the interaction with the wall is
 represented as an external force $F_w$ normal to the wall and  
decaying exponentially \cite{KWOK2, sullivan}, i.e.
\be\label{wallint}
F_{w}(\x)=\G_{w} \rho(\x) e^{- \left|\x-\x_{w}\right| / \xi} 
\ee
where $\x_{w}$ is a vector running along the wall location and $\xi$
the typical length-scale of the fluid-wall interaction. 
Equation (\ref{wallint}) has been previously used in literature by
using a slightly different LBE scheme to show how the wetting angle depends
on the ratio $\G_{w}/\G_{b}$ in presence of phase coexistence between
 vapor and liquid \cite{KWOK2}.  Here we want to study the opposite
 situation, i.e. the  effects of
 $\G_w$ when the thermodynamically stable bulk physics is governed by
 a single phase. The main result is that the presence of the wall may
 trigger a {local} phase coexistence inducing the formation of a
 less dense phase in the vicinity of the walls and an {apparent}
 slip of the bulk fluid velocity profile extrapolated at the wall location. 
%In figure
%\ref{fig1}, we show that also  the scheme here proposed gives a
%physically meaningful result for the  contact angle.  As expected, the contact angle
%decreases (the fluid is more wetting) upon increasing the attractive strength ($\G_{w}<0$) of
%the fluid-wall coupling.  According to our data, this decrease is
%close to linear and covers a significant range of contact angles.
%This curve can be regarded as a form of calibration of the mesoscopic
%LB models, much the same way the parameters of the Lennard-Jones
%potential are adjusted in molecular dynamics simulations.
%%%%%%%%%%%%%%%%%%%%%%%%%%%%%%%%%%%%%%%%%%%%%%%%%%%%%%%%%
\begin{figure}%[!t]
\begin{center}
\includegraphics[scale=.5]{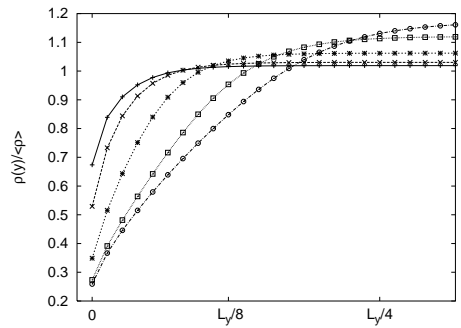}
 \caption{Rarefaction effects in the full-interaction case
   $\G_{w},\G_{b} \neq 0$. Density profiles are plotted as a function
   of the distance  from the wall, normalized to the channels height
   ($y/L_{y}$).  The wall interactions have been 
fixed assuming $\G_{w}=0.03$ and $\xi=2$. The
following values of $\G_{b}$ are considered: $\G_{b}=1.5$ ($+$), $\G_{b}=2.5$
($\times$), $\G_{b}=3.5$ ($\star$), $\G_{b}=3.9$ ($\square$), $\G_{b}=3.98$
($\circ$). We remind that the bulk phase transition is set at $\G^c_w
   = 4$. 
In all simulations we choose $L_{y}=80$ grid points for the height of
the channel. The volume averaged Knudsen is $Kn \sim 10^{-3}$, this 
would correspond to a channel of a few $\mu$m for liquid water.}
\label{fig3}
\end{center} 
\end{figure} 
Equations (\ref{1}-\ref{wallint}) have been numerically solved for
different values of the parameters $\G_b,\G_w$ and $\xi$ in a two
dimensional channel with periodic boundary conditions in the
stream-wise $x$ direction, being $y=0$ and $y=L_{y}$ the wall positions.
The sign of $\G_w$ is such to give a repulsive force for the liquid
particles at the wall.

The flow is driven by a constant pressure gradient in the $x$
direction $F_i = \delta_{i,x} \partial_x P_0$.
No-slip boundary conditions are used at the wall and for small
Knudsen numbers, i.e. in the large scale limit, the numerical
solutions have been checked against its weakly-inhomogeneous
macroscopic hydrodynamic limit, namely:
\begin{eqnarray}
&&\partial_t \rho + \partial_i (u_i\rho) = 0 \label{m1} \\
&&\rho \left[ \partial_{t} u_{i} + (u_{j} \partial_{j}) u_{i} \right] = 
   -\partial_{i} P 
+ \nu \partial_{j} (\rho \partial_{i} u_{j} + \rho \partial_{j} u_{i})
+ F_i
\nonumber \\
&& P=c_{s}^{2}\rho-V_{eff}(\rho) \label{m3}  \nonumber
\end{eqnarray}
where subscripts $i,j$ run over the two spatial dimensions.  
Above we  have $\nu=c^{2}_{s}(\tau-1/2)$ and $P$ is the total pressure
consisting of an ideal-gas contribution, $c_{s}^{2}\rho$, plus the
so-called excess pressure, $V_{eff}$,  due to potential-energy
interactions.  The expression of
$V_{eff}$ in terms of both $\G_{b}$ and $\G_{w}$ reads:
$$
V_{eff}(\rho) = \frac{1}{2}\G_b(1-\exp(-\rho))^2 +  \G_w \int_0^y ds
\rho(s)  \exp(-s/\xi).
$$
%%%%%%%%%%%%%%%%%%%%%%%%%%%%%%%%%%%%%%%%%%%%%%%%%%%%%%%%%
%
%\begin{figure}[!t]
%\begin{center}
%\includegraphics[scale=.5]{gwall.eps}
%\caption{Rarefaction effect as a function of the wall interactions for $\G_{b}=0$. We plot the ratio $\rho_{max}/\rho_{min}$ as a function of $G_{wall}$ for different values of the correlation length $\xi$: $\xi=1$ ($\circ$), $\xi=2$ ($\square$), $\xi=3$ ($\times$).}
%\label{fig2}
%\end{center} 
%\end{figure} %
%
%%%%%%%%%%%%%%%%%%%%%%%%%%%%%%%%%%%%%%%%%%%%%%%%%
Let us notice that the continuum equation (\ref{m1}) naturally
predicts the increase of the mass flow rate in presence of a density
profile which becomes more and more rarefied by approaching the wall
\cite{tretheway}.  Indeed, under stationary conditions, the continuity
equation in (\ref{m1}) reduces to $\partial_y (\rho u_y) = 0$, which,
because of the boundary conditions, implies $\rho u_y = 0$, i.e. $ u_y
= 0$ everywhere.  Thus, in  a homogeneous channel along the stream-wise
direction, the velocity $u_x$ satisfies the equation \be
\label{streamwise}
\nu \partial_y(\rho\partial_y u_x) = -\partial_x P_0.
\ee
In the new variable, 
 $y^{\prime}=y-H$, where  $H=L_{y}/2$, we may express the solution of  (\ref{streamwise})
as: 
\be
\label{ux}
 u_x(y^{\prime}) = -  \int_{y^{\prime}}^H
\frac{s \partial_x P_0 }{\nu
\rho(s)} ds.
\ee
Using (\ref{ux}) and assuming that density variations are concentrated
in a smaller layer 
of thickness $\delta$ near the wall,
 we can estimate the mass flow rate $Q_{eff}$ for small $\delta$ as:
\be 
\label{delta}
\frac{Q_{eff}}{Q_{pois}} = 1 + \frac{3}{2}\frac{\Delta
\rho_{w}}{\rho_w} \frac{\delta}{H} \ee where $Q_{pois}$ corresponds to
the Poiseuille rate $2 \partial_x P_0 H^3/3\nu$ valid for
incompressible flows with no-slip boundary conditions.  In equation
(\ref{delta}), the quantity $\Delta\rho_w$ is defined as the
difference between $\rho$ computed in the center of the channel and
$\rho_w$ computed at the wall. The effective slip length is then
usually defined in terms of the increment in the mass flow rate
\cite{brenner}: \be \label{slipdeltarho} \lambda_s \sim \delta
\frac{\Delta \rho_{w}}{\rho_w}.  \ee This is the best one can obtain
by using a purely continuum approach.  The added value of the
mesoscopic approach here proposed consists in the possibility to
directly compute the density profile, and its dependency on the
underlying wall-fluid and fluid-fluid physics.  To this purpose,  we
consider the momentum balance equation in (\ref{m1}) for the direction
normal to the wall, $i=y$. Since $u_y=0$, we simply obtain $\partial_y
P = 0$, i.e.  \be
\label{eqrho}
c^2_s \partial_y \rho - 2 \G_b (1-e^{-\rho})e^{-\rho}\partial_y \rho - \G_w \rho
e^{-y/\xi} = 0.
\ee
Let us first study the effects of the wall in  (\ref{eqrho}) 
by setting $\G_b=0$. One
can easily obtain $\log(\rho(y)/\rho_w) = \frac{\xi \G_w}{c^2_s}
(1-\exp(-y/\xi))$, which enables us to estimate $\Delta \rho_{w} =
\rho_w(\exp(\xi \G_w/c^2_s)-1)$. Using (\ref{slipdeltarho}), we obtain for
the effective slip-length:
\be
\label{est1}
\lambda_s \sim \xi e^{\xi \G_w/c^2_s} \,\,\, \ \ \ [\G_b = 0].  \ee 
%
%In figure
%\ref{fig2} we compare the solution of equation (\ref{eqrho}) for
%$\G_b=0$ and different values of $\xi$ (continuous lines) with respect
%to the numerical integration of the LBE equations
% (\ref{1}-\ref{wallint}): an excellent agreement is observed.\\ 
We now turn our attention to the non trivial interference between bulk
and wall physics whenever  $\G_b >0$.  Defining the bulk
pressure as: $P_{b} = c^2_s \rho - \frac{1}{2}\G_b(1-\exp(-\rho))^2$,
we can  rewrite equation (\ref{eqrho}) to highlight its physical
content as follows:
\be \log \left( \frac{\rho(y)}{\rho_{w}} \right) = \xi \G_w
(1-e^{-y/\xi})/\overline{\partial P_b/\partial \rho } \label{result}
\ee 
where the bulk effects appear only through the following term:
 \be
\overline{\frac{\partial P_b}{\partial \rho}} \equiv
\frac{1}{\log(\rho(y)/\rho_w)} \int_0^y \frac{\partial P_b}{\partial
\rho} \frac{d\rho}{\rho} \label{prho}.
\ee
%%%%%%%%%%%%%%%%%%%%%%%%%%%%%%%%%%%%%%%%%%%%%%%%%
Equation (\ref{result}) highlights two results.
First,  the effect of the bulk can always be interpreted as a
renormalization of the wall-fluid interaction by 
\be
\label{GR}
\G_w^R \equiv \G_w /\overline{\frac{\partial P_b}{\partial \rho} }.
\ee
%%%%%%%%%%%%%%%%%%%%%%%%%%%%%%%%%%%%%%%%%%%%%%%%%%%%%%%%%
\begin{figure}[!t]
\begin{center}
\includegraphics[scale=.5]{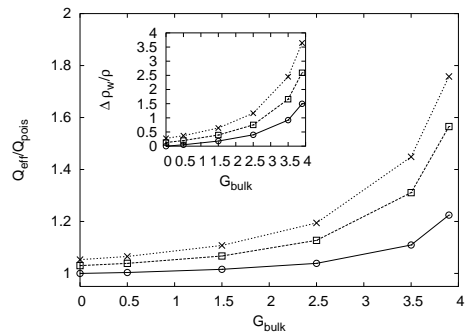}
\caption{Increase of the mass flow rate with the coupling strength
$\G_{b}$ of fluid-fluid bulk interactions. 
Fixing the wall correlation function $\xi=2$, we plot the mass flow rate ($Q_{eff}$)
normalized to its Poiseuille value ($Q_{pois}$) as a function of $\G_{b}$ for
different values of $\G_{wall}$: $\G_{wall}=0.0$ ($\circ$), $\G_{wall}=0.04$
($\square$), $\G_{wall}=0.08$ ($\times$). Inset: same as the main figure for $\Delta
\rho _{w} / \rho$.}
\label{fig4}
\end{center} 
\end{figure}
Second, as it is evident from (\ref{GR}), one must notice that near
the {bulk} critical point where $\partial P_b /\partial \rho
\rightarrow 0$,  the renormalizing effect can become unusually great. In other
words, the presence of the wall may locally push the system  toward a 
 phase transition even if the bulk physics it is far from the
 transition point.
As a result, the effective slip length in presence of both wall and
bulk non-ideal interactions  can be estimated as:
\be
\lambda_s \sim \xi \exp(\xi \G_w^R)
\label{deltaGR}
\ee In Fig.  \ref{fig3} we show $\rho(y)$ for different values of
$\G_b$ and $\G_w=0.03$, $\xi=2$ as obtained by numerically
integrating equations (\ref{1}-\ref{wallint}). The numerical
simulations have been carried out by keeping fixed the value of $\la
\rho \ra = \frac{1}{L_{y}} \int_0^{L_y}\rho(s) ds = \log(2)$. As one
can see, while $\G_b \rightarrow \G_c = 4$, the density difference
$\Delta \rho_{w}$ between the center of the channel and the wall
increases, as predicted by equation (\ref{eqrho}). Consequently, the
mass flow rate increases as shown in Fig.  \ref{fig4}. Let us notice
in the same figure 
that also with  $\G_w=0$, the wall initiates a small  rarefaction
effects due to the fact that fluid particles close to the boundary
 are attracted only by particles in the bulk of the channel. What
we showed here is that the combined actions of $\G_w$ and $\G_b
\rightarrow \G_b^c$ may strongly increase the formation of this less
dense region in the proximity of the surface.
%%%%%%%%%%%%%%%%%%%%%%%%%%%%%%%%%%%%%%%%%%%%%%%%%
For a quantitative  check, we have numerically integrated equations (\ref{eqrho}) and
(\ref{ux}) for a given value $\la \rho \ra = \log(2)$. The analytical
estimate for $\rho u_{x}$ is compared with the numerical results  in Fig. 
\ref{fig5}. This is a stringent test for our analytical interpretation. The result
is that the analytical estimate is able to capture the deviations from a pure
parabolic profile at approaching the wall region, where rarefaction effects are
present.
%%%%%%%%%%%%%%%%%%%%%%%%%%%%%%%%%%%%%%%%%%%%%%%%%%%%%%%%%
\begin{figure}[!t]
\begin{center}
\includegraphics[scale=.5]{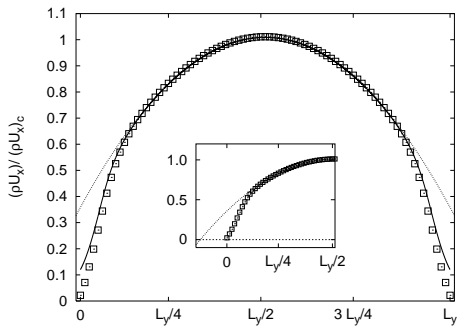}
\caption{Momentum profile as a function of the channel height. We plot the momentum
profile ($\rho u_{x}$) normalized to its center channel value ($(\rho u_{x})_{c}$)
as a function of the distance from the wall ($y$) normalized to the channel height
($L_{y}$). The results of numerical simulations ($\square$) with $\G_{b}=3.5$,
$\G_{w}=0.08$ and $\xi=2$ are compared with the analytical estimate (continuous
line) obtained solving equations (\ref{eqrho}) and (\ref{ux}). To highlight the
rarefaction effect, the parabolic fit in the center channel region (dotted line) is also plotted. Inset: estimate of the {apparent} slip length in the channel obtained the same
parabolic fit as in the main figure.}
\label{fig5}
\end{center} 
\end{figure} 
%%%%%%%%%%%%%%%%%%%%%%%%%%%%%%%%%%%%%%%%%%%%%%%%%
The crucial point in our analysis is that, even for very small $\G_w$,
large apparent slip can occur in the channel if $\G_b$ is close to its
critical value, i.e. the limit $\G_w \rightarrow 0$ and $\G_b
\rightarrow \G_b^c$ do {\it not} commute.  For example, let us
consider the case when $\G_w \sim \epsilon \ll
1$, $\xi \sim \epsilon$ and $\G_b \sim \G_b^c - \epsilon^3$, 
we obtain $ \overline{\frac{\partial P_b}{\partial \rho} } \sim \epsilon^3$ and
therefore, equation (\ref{deltaGR}) predicts that $\lambda_s \sim O(1)$
for $\epsilon \rightarrow 0$.  The wall effect, parametrized by $\G_w$
and $\xi$, can act as a catalyzer in producing large apparent slip.
Most of the results shown in Figs. (\ref{fig3}) and (\ref{fig4}) are
in close agreement with the MD numerical simulations 
 \cite{MD2,MD6,troian,barrat}. Our analysis points out
that, close to the wall, one can observe a ``local phase transition''
triggered by the presence of the wall itself.  In summary, we have
shown that a suitable form of the Lattice Boltzmann Equation can be
proposed in order to simulate apparent slip in microchannel.  Slip
boundary conditions arise spontaneously because, close to the wall, a
``gas'' layer is formed.  If the system is close to a state where
coexistence of different phases (liquid and gas) are thermodynamically
achievable, then, macroscopic slip effects can result.  We have shown that
for large scale separation, the model reduces to a continuum set of
hydrodynamical equations which explains the qualitative and
quantitative behavior of the mass flow rate in terms of the model
parameters, i.e. $\G_b$ and $\G_w$. 
%Comparison against experimental
%data can be investigated by choosing the ratio $\G_w/\G_b$ as a
%function of the contact angle.

\end{document}